\documentclass[prb,superscriptaddress,twocolumn]{revtex4-2}
\usepackage{graphicx}
\usepackage{times,amsmath}
\usepackage{epsfig}
\usepackage{color}
\usepackage{graphicx}% Include figure files
\usepackage{dcolumn}% Align table columns on decimal point
\usepackage{bm}% bold math
\usepackage{bookmark}
\usepackage{tabularx}% bold math
\usepackage{hyperref}
\usepackage{multirow}
\usepackage{array,mathtools,amssymb,booktabs,makecell}
\hypersetup{colorlinks=true, citecolor=blue, filecolor=blue, linkcolor=blue, urlcolor=blue}

\usepackage{float}
\usepackage{adjustbox}
\usepackage{upgreek}
\usepackage{soul}

\makeatletter
\let\saved@includegraphics\includegraphics
\AtBeginDocument{\let\includegraphics\saved@includegraphics}
\renewenvironment*{figure}{\@float{figure}}{\end@float}
\makeatother

\begin{document}

\title{\textcolor{black}{Lifshitz transition in correlated topological semimetals}}

%\title{Self-doping in strongly correlated topological semimetal}

%\title{Temperature dependent self-doping in strongly correlated topological semimetal}

\author{Byungkyun Kang}
\email[]{bkkang@utep.edu}
\affiliation{Department of Physics 500 W University Ave, The University of Texas at El Paso, El Paso, Texas 79968, USA}

\author{Myoung-Hwan Kim}
\affiliation{Department of Physics and Astronomy, Texas Tech University, Lubbock, Texas 79409, USA}

\author{Chul Hong Park}
\affiliation{Quantum Matter Core-Facility and Research Center of Dielectric and Advanced Matter Physics, Pusan National University, Busan 46240, Republic of Korea}

\author{Anderson Janotti}
\affiliation{Department of Materials Science and Engineering, University of Delaware, Newark, Delaware 19716, USA}

\author{Eunja Kim}
\affiliation{Department of Physics 500 W University Ave, The University of Texas at El Paso, El Paso, Texas 79968, USA}

\begin{abstract}
Topological quasiparticles, arising when the chemical potential is near the band crossing, are pivotal for the development of next-generation quantum devices. 
\textcolor{black}{They are expected to exist in half-Heusler correlated topological semimetals. However, the emergence of hole carriers, which alter the chemical potential away from the quadratic-band-touching points
is not yet understood.} 
Here, we investigated the electronic structure of \textcolor{black}{YPtBi and} GdPtBi through ab initio many-body perturbation GW theory combined with dynamical mean-field theory and revealed that \textcolor{black}{the correlation effects of 4$d$ or 4$f$ electrons can lead to the formation of hole carriers.}
\textcolor{black}{In YPtBi, the weakly correlated Y-4$d$ electrons constitute the topological bands, and the quadratic-band-touching point is at the Fermi level at high temperatures. At low temperatures, enhanced correlations of Y-4$d$ renormalize the topological bands, leading to the formation of hole pocket. In GdPtBi,} the \textcolor{black}{strongly correlated} Gd-4$f$ electrons form the Hubbard-like bands originate from self-energy effects associated with a topological singularity. These local bands encompass itinerant 4$f$ bands, which hybridize with topological bands to induce pronounced hole bands. 
This concerted effect reduces the hole doping, bringing the chemical potential closer to the quadratic-band-touching points as the temperature is lowered. 
The temperature-induced Lifshitz transition should be responsible for the large hole bands observed in \textcolor{black}{both topological semimetals} in angle-resolved photoemission spectroscopy measurements at \textcolor{black}{low temperatures.}
Our findings indicate that the integration of  correlated fermions within a topological framework can modulate the energy landscape of topological bands.
\end{abstract}

\maketitle

\textit{Introduction.} Topological semimetals and topological quasiparticles near Weyl points or quadratic-band-touching points have garnered significant attention due to their intriguing and unconventional physical properties, making them promising candidates for developing robust quantum devices~\cite{xu_prl2017,barry_science2016,lv_nphys2015,hu_apl2020,su_sciadv2017,kang_defect,yang_apl2019, yang_npjqm2022, daniel_ncom2020,pascal_prl2020}.
Research on Weyl fermions has primarily concentrated on weakly correlated electron systems. However, the electronic correlations present in \textcolor{black}{topological} systems are renowned for prompting novel electronic properties that are not typically observed in standard metals or insulators.
In this context, rare-earth-based \textcolor{black}{(including yttrium)} topological half-Heusler semimetals are suggested to display exotic physics, including tunable normal-state band inversion strength, superconducting pairing, magnetically ordered ground states, \textcolor{black}{unusual topological surface states,} anomalous Hall effects, heavy fermion behavior, quantum criticality, and Weyl fermions in strongly correlated electron systems~\cite{yasuyuki_sciadv2015,chandra_pnas2018,jie_prb2021,guo_ncom2018,max_nmat2016,suzuki_nphys2016,clemens_prb2020, eundeok_prb2016, sukhanov_prb2020, fisk_prl1991,guo_aip2018, mun_prb2013, ueland_prb2014, lin_nmat2010, canfield_jap1991,liu_ncomm2016,hosen_scirep2020,butch2011superconductivity,hyunsoo_sciadv2018}.

\begin{figure}[ht]
\centering
\includegraphics[width=0.5
\textwidth]{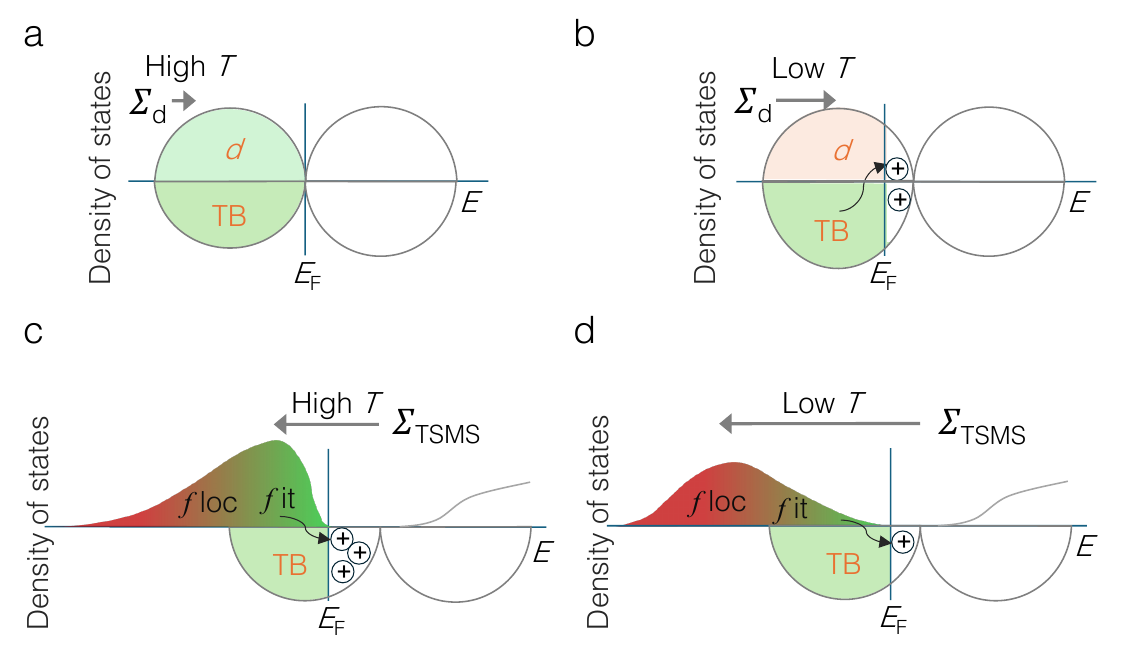}
\caption{\label{Fig_interplay}\
\textbf{Schematic diagram of the \textcolor{black}{impact of electron correlations on the formation of topological hole bands.}}
\textcolor{black}{\textbf{a}, In YPtBi, when the Y-4$d$ orbitals contribute the topological bands (TB), along with other orbitals, there is no significant self-energy of Y-4$d$ at high temperature. In this case, the quadratic-band-touching points (QTP) are at the chemical potential ($E_\text{F}$).
\textbf{b}, In YPtBi, the enhanced self-energy of Y-4$d$ at low temperature increased energy level of the TB including Y-4$d$ bands. This increase density of states of the occupied TB leading to the formation of hole bands.} 
\textbf{c},
In GdPtBi, the topological singularity-induced Mott-like self-energy (TSMS, horizontal arrow) at the QTP gives rise to broadened Hubbard-like bands, which comprise the 4$f$ local bands ($f$ loc) and the 4$f$ itinerant bands ($f$ it). These bands act as hole dopants for the TB, shifting the $E_\text{F}$ downward from the QTP. At high temperatures (High $T$), TSMS pushes the Hubbard-like bands away from the QTP, with an overlap between $f$ it and TB, which leads to hole doping and $E_\text{F}$ shifting.
\textbf{d},
In comparison to High $T$, at Low temperature (Low $T$), the TSMS strengthens, pushing the Hubbard-like bands further from the QTP. This reduces the overlap between $f$ it and TB, diminishes hole doping of the TB, and leads to a smaller shift of $E_\text{F}$ away from the QTP.
}
\end{figure}

In RPtBi (R = rare earth element \textcolor{black}{and Y}), the topology of the Fermi surface is highly sensitive to the choice of R, significantly influencing exotic quantum phenomena~\cite{canfield_jap1991,stanislav_nmat2010,eundeok_mrs2022}. One proposed scenario for the impact of R-4$f$ electrons on topological bands at the Fermi level involves the Kondo effect. It is suggested that electronic bands near the Fermi level may become heavily renormalized due to the strong Kondo coupling between R-4$f$ electrons and conduction band states below the Kondo temperature, which significantly affects the topologically non-trivial state in SmB$_6$~\cite{maxim_prl2010,steven_prb2013,maxim_arcmp2016}.
For the Kondo effect to occur, $f$ electrons must be close to the Fermi level. 
However, the varying energy levels of quadratic-band-touching points in RPtBi are not explained by the Kondo effect alone. 
Specifically, angle-resolved photoemission spectroscopy (ARPES) measurements of GdPtBi at 15 K, which features a strongly localized 4$f$ electron shell far from the Fermi level and large hole bands, have deduced quadratic-band-touching points at 0.4 eV above the Fermi level, with no indication of a saddle point near the Fermi level~\cite{chang_prb2011}. 
In contrast, longitudinal magnetoresistance experiments~\cite{max_nmat2016} indicate the existence of the chiral anomaly in p-type GdPtBi samples at 2.5 K. This observation relies on a model featuring quadratic-band-touching points near the Fermi level. When a magnetic field of $B$ = 6 T is applied, the Zeeman splitting positions the chemical potential between saddle points, which is essential for the occurrence of the chiral anomaly.
The band structure calculations of GdPtBi, using density functional theory combined with Hubbard $U$ (DFT+$U$), showed that the quadratic-band-touching points are located precisely at the Fermi level~\cite{max_nmat2016,suzuki_nphys2016}.
In the research, the splitting of these quadratic bands remains below 0.1 eV, when a magnetic field of $B$ = 8 T is applied~\cite{suzuki_nphys2016}.
This indicates that the quadratic-band-touching points identified at 0.4 eV by the ARPES~\cite{chang_prb2011} are unlikely to develop into a chemical potential between saddle points through the Zeeman splitting.
Crystallographic defects in the topological semimetal have the potential to induce charge doping and a hole band~\cite{kang_defect,khalid2022defect}. However, evidence for the formation of a large hole band, as measured by ARPES, remains elusive in GdPtBi because it would require a high concentration of defects. 
\textcolor{black}{Besides of the R-4$f$, ARPES measurements of YPtBi reveal that the quadratic-band-touching point lies above the Fermi level and a large hole band is present~\cite{liu_ncomm2016,hosen_scirep2020}, indicating a significant impact of Y-4$d$ electrons on the topological bands structure.}
These findings suggest that the electronic structure of RPtBi near the Fermi level is unclear, indicating that an unknown many-body effect associated with temperature impacts the system.

\begin{figure}[ht]
\centering
\includegraphics[width=0.5
\textwidth]{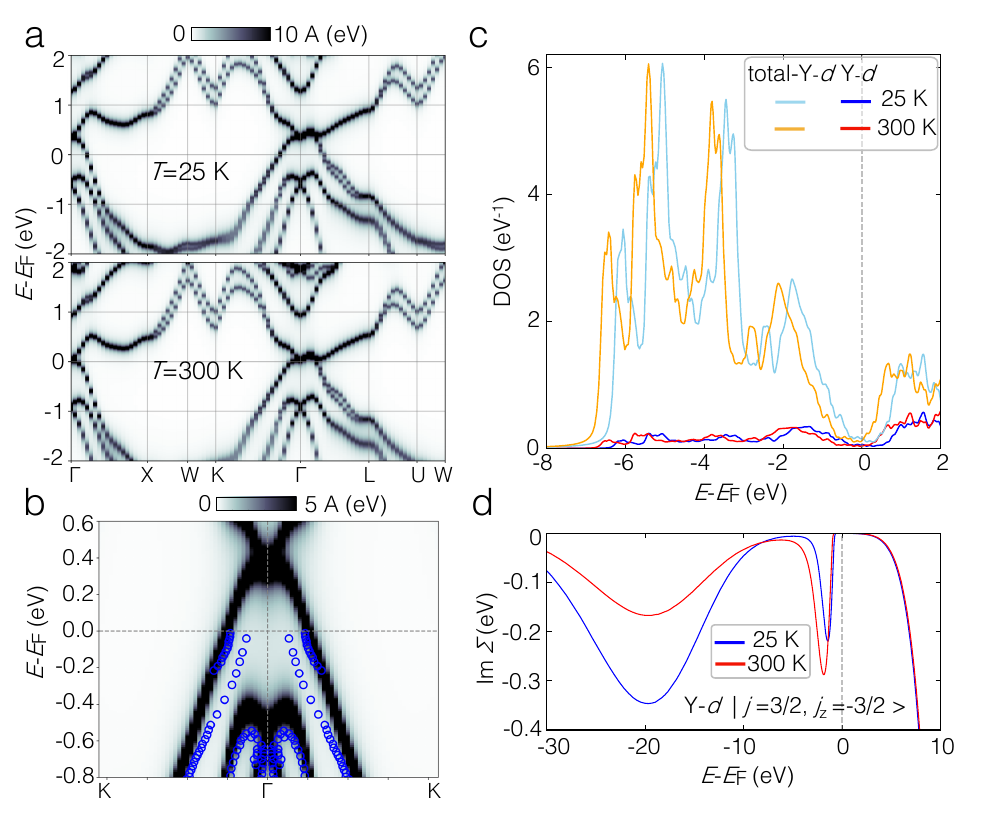}
\caption{\label{Fig_yptbi}\
\textbf{\textcolor{black}{Electronic structure of YPtBi.}}
\textcolor{black}{
\textbf{a}, Spectral functions at $T$ = 25 K (upper panel) and $T$ = 300 K (lower panel).
\textbf{b}, Calculated spectral function at $T$ = 25 K, along the K-$\Gamma$-K high symmetry line. ARPES measurements at $T$ = 20 K~\cite{liu_ncomm2016} are denoted by blue dots.
\textbf{c}, Density of states of Y-4$d$ projected and total subtracted by Y-4$d$ at 25 and 300 K. 
\textbf{d}, The imaginary part of the self-energy of the Y-4$d$ state of the $|j= 3/2,j_{z}= -3/2\rangle$.
}
}
\end{figure}

Recent studies in quantum many-body systems have shown that the dynamic properties of R-4$f$ electrons vary considerably depending on the specific class of R-based materials. In the case of the superconductor NdNiO$_{2}$, Nd-4$f$ electrons are strongly localized, resulting in the Kondo effect~\cite{kang_ndnio2}. In contrast, Ce-4$f$ electrons in Ce films exhibit itinerancy, forming dispersive quasiparticle bands~\cite{wu_ncom2021}. Research by Park et al.~\cite{park_pnas2008} illustrates that the single 4$f$ electron in cerium within CeRhIn$_{5}$ simultaneously contributes to magnetism, a characteristic of localization, and superconductivity, which requires itinerancy.
According to our recent study \cite{kang_tsme}, the screening suppression of R-4$f$ at topological singular points in RPtBi and RAlGe prompts a robust topological singularity-induced Mott-like self-energy (TSMS).
In this study, we discovered that GdPtBi exhibits an interplay of the dual nature of Gd-4$f$ and TSMS. As illustrated in Fig.~\ref{Fig_interplay}\textcolor{black}{ c and d}, the TSMS gives rise to Gd-4$f$ Hubbard-like local bands, which incorporate the Gd-4$f$ itinerant bands. The latter hybridizes with topological bands, leading to topological hole bands. The Hubbard-like bands are pushed down from the quadratic-band-touching points by TSMS, which becomes more pronounced at lower temperatures. This interplay leads to a Lifshitz transition in GdPtBi. 
\textcolor{black}{In contrast, as shown in Fig.~\ref{Fig_interplay}a and b, the Y-4$d$ orbitals contribute to the formation of topological bands in YPtBi, along with other orbitals. The enhanced self-energy of the Y-4$d$ at low temperatures leads to the emergence of topological hole bands.}

To account for the interplay \textcolor{black}{between electron correlation and topological bands}, we ascertain the electronic structure of GdPtBi \textcolor{black}{, YPtBi, }and YbPtBi using the ab-initio linearized quasiparticle self-consistent GW combined with the dynamical mean field theory (LQSGW+DMFT) approach~\cite{choi2019comdmft}, which is based on a simplified full GW+EDMFT approach~\cite{sun2002extended,biermann_prl2003,nilsson2017multitier,kangfgwedmft,kang_fese}. 
This method allows us to precisely determine the temperature-dependent electronic structure of strongly correlated materials without any subjective biases. While we adopt the experimental lattice parameters~\cite{martin_jssc2002}, we explicitly calculate all other parameters, including the double-counting energy and the Coulomb interaction tensor. 
\textcolor{black}{Analytical continuation from the imaginary to the real frequency axis was carried out for the self-energy using the maximum entropy method~\cite{jarrell_physrep1996}. The validity of this approach in our context is substantiated in Supplementary Figure~2.}
Further details can be found in the Supplementary Methods section. The LQSGW+DMFT approach has been applied to study many-body effects, including Kondo effects~\cite{byung_usbte,kang2023dual,byung_ute2,kang_ndnio2,kang_ute2fs}, Hund and Mott physics~\cite{kang_nqm2023,kang_nio}, and interplay between strong electronic correlations and nontrivial topology~\cite{kim_prl2024}.

\textcolor{black}{\textit{Temperature-induced Lifshitz transition in YPtBi.}} 
\textcolor{black}{
Figure~\ref{Fig_yptbi}a shows the calculated spectral functions of YPtBi. The presence of quadratic-band-touching points near the Fermi level at the $\Gamma$ point is apparent at both $T = 300$ K and $T = 25$ K. At 300 K, these band-touching points are located at the Fermi level, whereas at 25 K, they are shifted to approximately 0.4 eV above the Fermi level. In Figure~\ref{Fig_yptbi}b, the spectral function along the K-$\Gamma$-K high-symmetry direction at 25 K is compared with ARPES measurements performed at 20 K~\cite{liu_ncomm2016}. The ARPES data display a strong, disconnected linear feature with a weaker linear branch near the Fermi level, while our calculations reveal a prominent single linear dispersion intersecting the Fermi level. Notably, both ARPES measurements and theoretical calculations demonstrate the presence of quadratic-band-touching points above the Fermi level.
}

\textcolor{black}{
The Y-4$d$ electrons in YPtBi exhibit weak electronic correlations, as evidenced by the comparatively small value of the static on-site Coulomb interaction (obtained using cRPA; see Supplementary Figure 1), $U_{\textrm{C}}(\omega=0) = 1.9$ eV, and the large electronic bandwidth of approximately 7 eV (Fig.\ref{Fig_yptbi}c). Consequently, Y-4$d$ electrons contribute to the topological bands together with other electronic states in YPtBi, as illustrated in Fig.\ref{Fig_yptbi}c. These findings suggest that the Y-4$d$ electrons in YPtBi may be described as forming an electron gas, with their correlation effects potentially exhibiting temperature dependence\cite{kulveer_jpcm2020}.
}

\textcolor{black}{
In YPtBi, the self-energy associated with the Y-4$d$ states, which is centered below the topological bands at approximately -20 eV, exhibits a marked increase as the temperature is reduced from 300 K to 25 K, indicative of enhanced electron correlations within the Y-4$d$ orbitals at low temperatures, as shown in Fig.~\ref{Fig_yptbi}d. This enhancement in self-energy at 25 K elevates the energy level of the Y-4$d$ electrons. Furthermore, the strong hybridization between the Y-4$d$ electrons and those forming the topological bands leads to an upward shift in the energy levels of the topological bands, which in turn results in a lowering of the chemical potential and an increased density of states below the Fermi level.
The enhancement of the self-energy for Y-4$d$ states facilitates charge transfer from the Y-4$d$ orbitals to the topological bands (total states minus Y-4$d$ contribution). Although this charge transfer is relatively small, it is evident in the calculated Y-4$d$ occupancy, which decreases from 1.15 at 300 K to 1.09 at 25 K. Additionally, the peak in the (total - Y-4$d$) DOS shifts to a higher intensity near -1.8 eV at 25 K, compared to the corresponding peak near -2.2 eV at 300 K, as illustrated in Fig.~\ref{Fig_yptbi}c. These findings indicate that the increase in DOS within the topological bands (total - Y-4$d$) at lower temperatures predominantly contributes to effective hole doping.
}

\begin{figure}[ht]
\centering
\includegraphics[width=0.5
\textwidth]{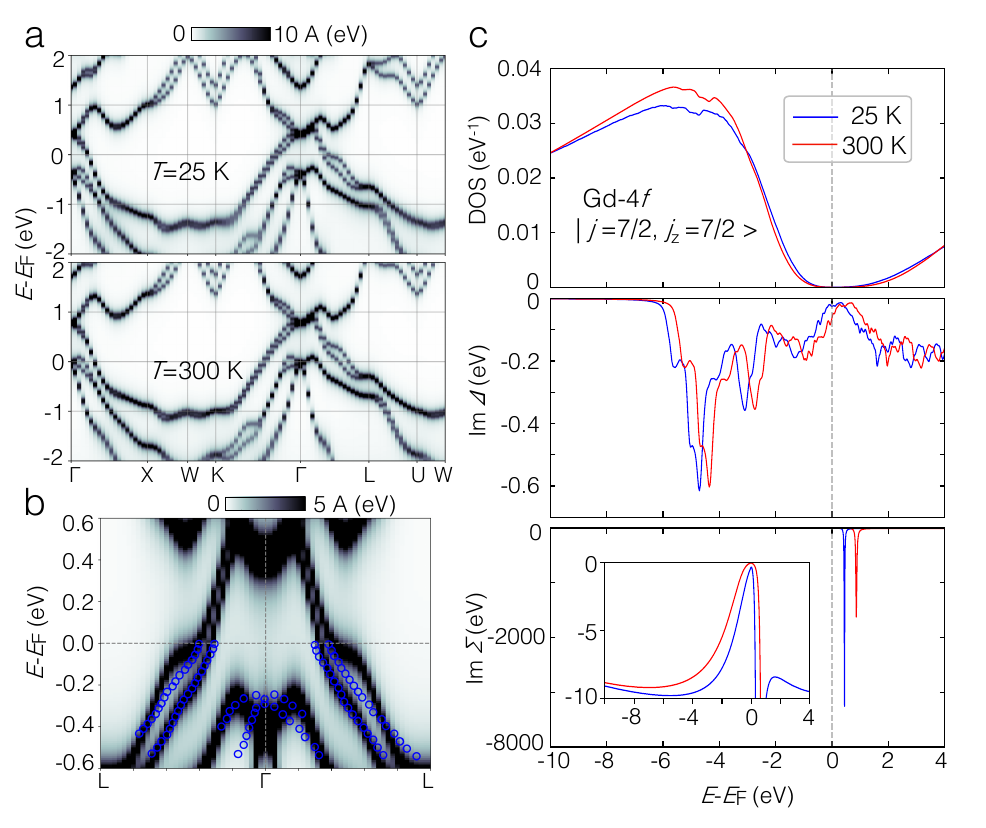}
\caption{\label{Fig_gdptbi}\
\textbf{Electronic structure of GdPtBi.}
\textbf{a}, Spectral functions at $T$ = 25 K (upper panel) and $T$ = 300 K (lower panel).
\textbf{b}, Calculated spectral function at $T$ = 25 K, along the L-$\Gamma$-L high symmetry line. ARPES measurements at $T$ = 15 K~\cite{chang_prb2011} are denoted by blue dots.
\textbf{c}, Projected DOS, the imaginary part of the hybridization function ($\Delta$) and imaginary part of the self-energy of the Gd-4$f$ state of the $|j= 7/2,j_{z}= 7/2\rangle$.
}
\end{figure}

\textit{Temperature-induced Lifshitz transition in GdPtBi.}
Figure~\ref{Fig_gdptbi}a presents the calculated spectral functions of GdPtBi. The quadratic-band-touching points near the Fermi level at the $\Gamma$ point are evident at both $T$ = 300 K and $T$ = 25 K. At 300 K, the quadratic-band-touching points are observed at $\sim$0.8 eV, shifting to $\sim$0.4 eV at 25 K.
The evolution of the Fermi surface with temperature is reminiscent of the temperature-induced Lifshitz transition in the topological insulator ZrTe$_5$~\cite{yan_nacomm2017}.
The spectral function at 25 K along the L-$\Gamma$-L high-symmetry line is compared with ARPES measurements at 15 K~\cite{chang_prb2011} in Fig.~\ref{Fig_gdptbi}b. Both the theoretical and experimental results exhibit two linear lines crossing the Fermi level of the hole bands. Notably, the approximate band crossing point of 0.4 eV, deduced from the ARPES measurements~\cite{chang_prb2011}, aligns well with the calculated quadratic-band-touching points in the spectral function.

\begin{figure}[ht]
\centering
\includegraphics[width=0.5
\textwidth]{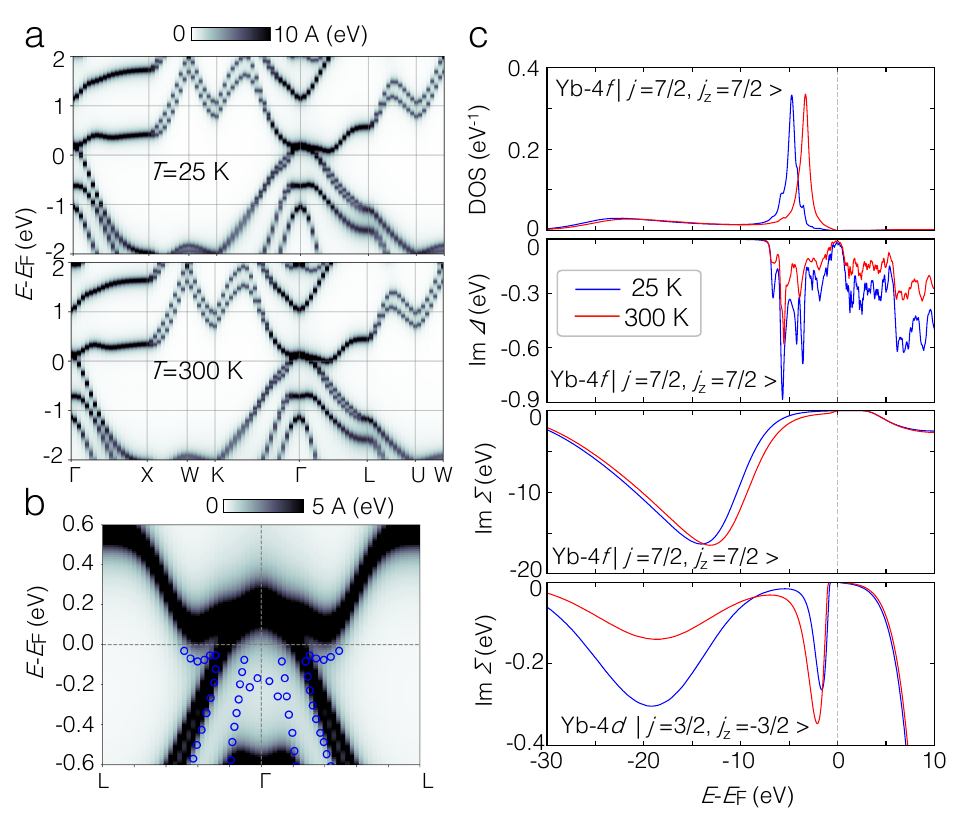}
\caption{\label{Fig_ybptbi}\
\textbf{Electronic structure of YbPtBi.}
\textbf{a}, Spectral functions at $T$ = 25 K (upper panel) and $T$ = 300 K (lower panel).
\textbf{b}, Calculated spectral function at $T$ = 25 K, along the L-$\Gamma$-L high symmetry line. ARPES measurements at $T$ = 20 K~\cite{guo_ncom2018} are denoted by blue dots.
\textbf{c}, Projected DOS, the imaginary part of the hybridization function ($\Delta$) and imaginary part of the self-energy of the Yb-4$f$ state of the $|j= 7/2,j_{z}= 7/2\rangle$.
}
\end{figure}

The temperature-induced Lifshitz transition in GdPtBi is attributed to the interplay of the dual nature of Gd-4$f$ electrons with TSMS. The calculated static $U_{\textrm{C}}(\omega=0)$ of Gd-4$f$ in GdPtBi is 3.4 eV.
\textcolor{black}{In comparison to the Mott-like self-energy of the Pr-4$f$ in PrPtBi, which is $\sim$-8 eV and is attributed to a larger on-site Coulomb interaction $U_{\textrm{C}}(\omega=0) = 6.3$ eV for Pr-4$f$, as reported in Ref.~\cite{kang_tsme}, GdPtBi exhibits a substantially greater self-energy for the Gd-4$f$, on the order of $\sim$-80 eV, despite having a smaller $U_{\textrm{C}}(\omega=0)$. This indicates a distinct correlation mechanism underlying the electronic structure of the Gd-4$f$ states in GdPtBi compared to the Pr-4$f$ states in PrPtBi.}
Therefore, the divergent self-energy peak observed, as depicted in Fig.~\ref{Fig_gdptbi}c, should be interpreted as TSMS. 
This arises from the concerted effect between R-4$f$ quasiparticles and the topological singularity, as demonstrated in our work~\cite{kang_tsme}. Gd-4$f$, being subject to intermediate Coulomb repulsion, has a higher probability of forming quasiparticles near the Fermi level at lower temperatures \cite{antonio_rmp1996,kang_ndnio2}. This leads to increased screening suppression at the quadratic-band-touching points near the Fermi level, resulting in enhanced TSMS at 25 K, as illustrated in the self-energy plot in Fig.~\ref{Fig_gdptbi}c.
Figure~\ref{Fig_gdptbi}c illustrates the density of states (DOS), hybridization function, and self-energy of the Gd-4$f$ state of the $|j= 7/2, j_{z}= 7/2\rangle$.
The strong TSMS induces lower Hubbard-like incoherent bands in the DOS, a hallmark of Mott physics that indicates the localized behavior of correlated electrons.
Small peaks (or shoulders) in the DOS appear at approximately -4.0 eV and -2.3 eV at 50 K, shifting to -4.4 eV and -2.7 eV at 25 K, respectively.
\textcolor{black}{There is no sizable peaks in the self-energy around these energy level as shown in inset in Fig.3c.}
These peaks\textcolor{black}{, therefore, can be associated with the major peaks in} the Gd-4$f$ hybridization functions \textcolor{black}{at the corresponding energy levels}, suggesting an interaction between the Gd-4$f$ state and the topological bands. These findings imply that the Gd-4$f$ state in GdPtBi possesses a dual character, exhibiting both localized and itinerant electron behavior.

As shown in Fig.~\ref{Fig_gdptbi}c, the hybridization function maintains a consistent magnitude across both temperatures, demonstrating that Gd-4$f$ electrons remain fully hybridized with the topological bands irrespective of temperature. Consequently, Gd-4$f$ electrons effectively introduce hole doping to the topological bands. This interaction pushes the chemical potential down from the quadratic-band-touching points, consistently observed at both temperatures.
The extent of self-doping is influenced by the strength of TSMS. At 25 K, compared to 300 K, an increase in TSMS causes the Gd-4$f$ Hubbard-like bands to move further from the quadratic-band-touching points. This shift reduces the Gd-4$f$ DOS between -6 eV and the quadratic-band-touching points, where hybridization occurs, and subsequently decreases the hole doping effect on the topological bands, adjusting the chemical potential closer to the quadratic-band-touching points. 
This behavior contrasts with that observed in LaPtBi, where the quadratic-band-touching points align with the Fermi level in the absence of 4$f$ electrons~\cite{kang_tsme}. Moreover, this is different compared to the temperature-dependent Nd-4$f$ bands, as they do not affect the chemical potential of NdNiO$_2$~\cite{kang_ndnio2}. The degree of self-hole doping, or the shift in the chemical potential of the topological bands, is dependent on the extent to which the Gd-4$f$ Hubbard-like bands overlap with the topological bands.

\textit{\textcolor{black}{Lifshitz transition is absent for Yb-4$f$ and present for Yb-4$d$ in YbPtBi.}}
Figure~\ref{Fig_ybptbi}a presents the calculated spectral functions of YbPtBi. It shows quadratic-band-touching points at the $\Gamma$ point, positioned in the vicinity of the Fermi level, in contrast to those in GdPtBi. Unlike GdPtBi \textcolor{black}{and YPtBi}, YbPtBi exhibit a \textcolor{black}{small} chemical potential shift when lowering the temperature from 300 K to 25 K. The spectral function at 25 K, along the L-$\Gamma$-L high symmetry line, is compared with ARPES measurements taken at 20 K~\cite{guo_ncom2018} in Fig.~\ref{Fig_ybptbi}b. Both the theoretical and experimental results manifest two hole pockets crossing the Fermi level and doubly degenerate $\Uplambda_{6}$ electron pockets near the Fermi level. 
Our results agree with DFT simulations in which Yb-4$f$ electrons are treated as core states~\cite{mun_prb2015,guo_ncom2018}.
This indicates that the \textcolor{black}{strong} many-body effect observed in GdPtBi is absent in YbPtBi, and the 4$f$ electrons in YbPtBi do not affect the topological bands near the Fermi level.

\begin{figure}[ht]
\centering
\includegraphics[width=0.5
\textwidth]{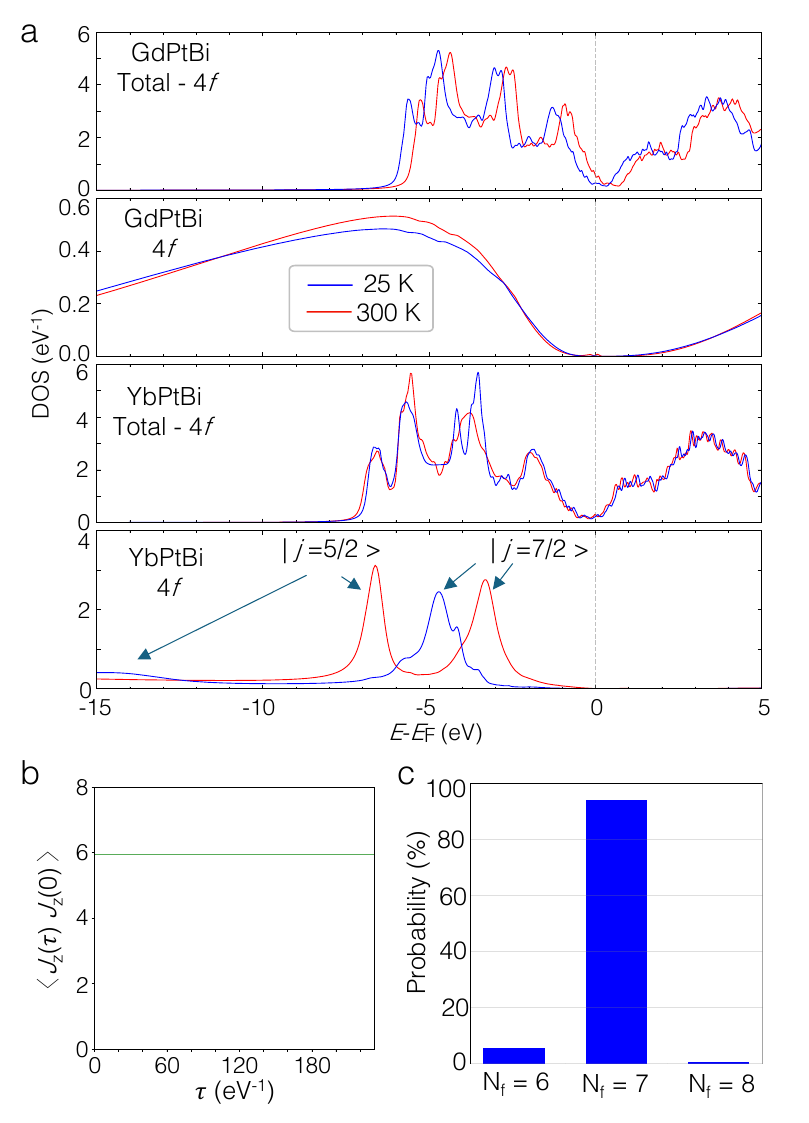}
\caption{\label{Fig_compare}\
\textbf{Comparison of electronic structures.}
\textbf{a}, The calculated DOS of GdPtBi: total DOS minus the 4$f$ contribution (first panel) and the 4$f$ contribution only (second panel) at 25 K and 300 K. Similarly, the calculated DOS of YbPtBi: total DOS minus the 4$f$ contribution (third panel) and the 4$f$ contribution only (fourth panel) at 25 K and 300 K. 
\textbf{b}, The local spin moment correlation function in the imaginary time $\tau$ of the Gd-4$f$ in GdPtBi at 25 K.
\textbf{c}, Valence histograms of the Gd-4$f$ in GdPtBi at 25 K.
}
\end{figure}

Figure~\ref{Fig_ybptbi}c shows the DOS, hybridization function, and self-energy of the Yb-4$f$ state of the $|j= 7/2,j_{z}= 7/2\rangle$. The fully filled Yb-4$f$ states are positioned far below the Fermi level in the DOS. However, small peaks at $\sim$-4 eV and $\sim$-6 eV in both the DOS and hybridization functions at 25 K indicates that the Yb-4$f$ states also hybridized with topological bands.
In contrast to GdPtBi, there is no TSMS in YbPtBi. 
Although cooling causes the Yb-4$f$ state to further distance themselves from the Fermi level, YbPtBi does not demonstrate \textcolor{black}{significant} temperature-induced changes in topological bands from -2 eV to 2 eV, as shown in Fig.~\ref{Fig_ybptbi}a. This is because the interaction between the 4$f$ and topological bands is too localized to induce the Lifshitz \textcolor{black}{transition.} \textcolor{black}{However, the small chemical potential shift is attributed to the enhanced self-energy of Yb-4$d$ (see the bottom panel of Fig.~\ref{Fig_ybptbi}c), similarly to YPtBi.}

\textit{Impact of topological singularity-induced Mott-like self-energy.}
The recent discovery of the TSMS in RPtBi, as reported by Kang et al.~\cite{kang_tsme}, sheds light on its significance in strongly correlated topological semimetals. To further elucidate the impact of the TSMS, a comparative analysis was conducted on the DOS of GdPtBi with YbPtBi, the latter of which lacks the TSMS. As shown in Fig.~\ref{Fig_compare},
the Yb-4$f$ DOS in YbPtBi exhibits local peaks centered at $\sim-3.3$ eV for the $j = 7/2$ multiplet and at $\sim-6.7$ eV for the $j = 5/2$ multiplet at 300 K. 
The $j = 7/2$ multiplet states are predominantly hybridized with topological bands (total - 4$f$ DOS) at $\sim -4.0$ eV. This is evident from the smaller DOS of the topological bands compared to that at 25 K, where the $j = 7/2$ multiplet states have shifted towards lower energy.
Due to the shift at 25 K, the $j = 7/2$ multiplet states hybridize at $\sim-6.0$ eV, resulting in a smaller DOS for the topological bands at $\sim-6.0$ eV compared to that at 300 K.
Although these temperature-dependent hybridizations alter the topological bands locally, they do not change the energy levels of the topological bands. This is attributed to the strongly localized DOS of Yb-4$f$ electrons, which can be treated as core electrons~\cite{mun_prb2015}.

However, the character of Gd-4$f$ in GdPtBi differs from the localized Yb-4$f$ in YbPtBi. 
The local Green's function in DMFT~\cite{book_dmft} is expressed as ${G}_{loc}^{-1}(i\omega_n) = i\omega_n + \mu - {\mathit{\Delta}}(i\omega_n) - \mathit{\Sigma}(i\omega_n)$ on the imaginary frequency, where $\mu$ is the chemical potential, $\mathit{\Delta}(i\omega_n)$ represents the hybridization function, and $\mathit{\Sigma}(i\omega_n)$ is the impurity self-energy.
In GdPtBi, self-energy is governed by TSMS. At low temperatures, Gd-4$f$ quasiparticles are highly formed at $\mu$ if there is no TSMS. 
When $\mu$ is at the quadratic-band-touching points, due to the incompatibility between the 4$f$ quasiparticle and the topological singular point at the same energy level~\cite{kang_tsme}, a diverging TSMS appears at $\mu$, leading the quasiparticles to significantly contribute to the Hubbard-like Gd-4$f$ spectral function.
Then, the $\mathit{\Delta}(i\omega_n)$, arising from the dual nature, affects the ${G}_{\text{loc}}(i\omega_n)$ by shifting $\mu$ downward, away from the quadratic-band-touching points.
The changing $\mu$, instead of altering ${G}_{loc}(i\omega_n)$ locally as in YbPtBi, is attributed to the strong TSMS, which leads to the incoherent broadening of the Hubbard-like Gd-4$f$ DOS. This hybridizes with most of the topological bands, unlike YbPtBi, as shown in Fig.~\ref{Fig_compare}.
Although the TSMS-induced Gd-4$f$ DOS broadens from the chemical potential to beyond -15 eV, Gd-4$f$ is strongly localized, i.e., Hubbard-like.
The local spin moment correlation function, $\chi_{J_Z}(\tau)=\langle J_z(\tau)J_z(0)\rangle$, is presented in Fig.~\ref{Fig_compare}b. This correlation function can be used to evaluate the extent of magnetic moment localization\cite{belozerow_prb2023}. The $\chi_{J_Z}(\tau)$ of Gd-4$f$ in GdPtBi is found to remain constant with an increase in $\tau$, suggesting the presence of strong local magnetic moments. 
This is reminiscent of the strong local spin moment found in the van der Waals 2D Mott-insulator FeSe~\cite{kang_fese}, \textcolor{black}{in the periodic Anderson model during its metal–insulator transition~\cite{amaricci_epl2017}, in iron and nickel at ambient and high pressure~\cite{hausoel_ncomm2017}, in heterogeneous Ta-dichalcogenide bilayers~\cite{lorenzo_ncom2023}, in strongly correlated Hund metals~\cite{stadler_allphy2019}, and in Fe-based superconductors~\cite{toschi_prb2012}.}
Additionally, as shown in Fig.~\ref{Fig_compare}c, the valence histogram of the Gd-4$f$ in GdPtBi exhibits subtle valence fluctuations, indicating that the local character of Gd-4$f$ dominates.

\textit{Conclusion.}
We show that the quadratic-band-touching points in GdPtBi are located at $\sim$0.4 eV above the Fermi level at 25 K, which agrees with the metallic Fermi surface measured by ARPES at 15 K~\cite{chang_prb2011}.
We also show that the chemical potential shifts from $\sim$0.8 eV to $\sim$0.4 eV, moving closer to the quadratic-band-touching points as the temperature decreases from 300 K to 25 K, due to enhanced TSMS at lower temperatures. 
This indicates that further cooling should draw the chemical potential nearer to the quadratic-band-touching points, aligning the Fermi level between the saddle points as a result of Zeeman splitting when a magnetic field is applied. 
Therefore, the temperature-induced Lifshitz transition should play a crucial role in the emergence of the chiral anomaly in p-type GdPtBi samples, as measured at $T$ = 2.5 K~\cite{max_nmat2016}.

In summary, 
\textcolor{black}{ in the weakly correlated topological semimetal YPtBi, the observed hole bands are attributed to a temperature-dependent Lifshitz transition resulting from enhanced electron correlation of Y-4$d$ electrons.}
\textcolor{black}{In the strongly correlated topological semimetal GdPtBi,} the Gd-4$f$ electrons, subjected to TSMS at quadratic-band-touching points, evolve into broadened Hubbard-like bands. These bands exhibit a dual character, comprising both localized and itinerant features. The itinerant bands interact with the topological bands, resulting in hole doping. The degree of this self-doping is influenced by the strength of the temperature-dependent TSMS, which leads to a temperature-induced Lifshitz transition.
The conventional understanding suggests that holes emerge due to intrinsic defects, and the Kondo hybridization involving R-4$f$ modifies low-energy physics, thus affecting topological quasiparticles. These perspectives have been limited to modulating correlated topological materials. 
However, our study introduces an intrinsic mechanism of self-doping based on the interaction between topological bands and correlated electrons, which influences exotic quantum phenomena in these materials.
\textcolor{black}{We propose two distinct mechanisms for Lifshitz transitions arising from the interplay between electronic correlations and topological band structures. These mechanisms may also be relevant to Lifshitz transitions observed in other topological materials, such as ZrTe$_2$\cite{yan_nacomm2017} and WTe$_2$\cite{yun_prl2015}. Moreover, our results may shed light on the origin of the Dirac node located below the Fermi level in Ce-intercalated graphene~\cite{jinwoong_nanolett2018}.}
This advancement should be instrumental in the design of topological superconductors and quantum computing devices within strongly correlated topological systems.

%\subsection*{Methods}
%\noindent \textbf{LQSGW+DMFT calculation\\}
%We adopted experimental lattice constants of space group $I4_{1}md$ (No. 109) (PrAlGe: $a=$ 4.253, $c=$ 14.641 $\textrm{\AA}$)~\cite{gladyshevskii_jac2000}, (LaAlGe: $a=$ 4.336, $c=$ 14.828 $\textrm{\AA}$)~\cite{arnold_inorgchem1991}, and space group $F\bar{4}3m$ (No. 216) (HoPtBi: $a=$ 6.631 $\textrm{\AA}$), (PrPtBi: $a=$ 6.78 $\textrm{\AA}$), (LaPtBi: $a=$ 6.829 $\textrm{\AA}$)~\cite{martin_jssc2002} (see Fig. S1).
%We employed the $ab$-$initio$ linearized quasiparticle self-consistent GW (LQSGW) method combined with dynamical mean field theory (DMFT)~\cite{choi2019comdmft}. This LQSGW+DMFT is based on a simplified full GW+DMFT approach~\cite{sun2002extended, biermann_prl2003, nilsson2017multitier, kangfgwedmft, kang_fese}. The electronic structure was calculated using LQSGW approaches~\cite{kutepov2012electronic, kutepov2017linearized}, and the local part of the GW self-energy was diagrammatically corrected within DMFT~\cite{georges1996dynamical, metzner1989correlated, georges1992hubbard}. We also explicitly calculated the double-counting energy and dynamically screened Coulomb interaction tensor (see Fig. S2). Local self-energies for R-4$f$ and R-5$d$ were obtained by solving impurity models, with spin-orbital coupling included in all calculations. The self-energy is computed along an imaginary frequency axis and then analytically continued through a maximum entropy method~\cite{jarrell_physrep1996}. Please see the details of this method in the Supplementary information.

\section*{Acknowledgments}  
We acknowledge the High Performance Computing Center (HPCC) at Texas Tech University for providing computational resources that have contributed to the research results reported within this paper.
C.H. Park acknowledges the support by the National Research Foundation of Korea (NRF) grant (Grant No. NRF-2022R1A2C1005548) and the Ministry of Education (grant No. 2021R1A6C101A429).

\bigskip
\textbf{Competing Interests} The authors declare no competing interests.

\bigskip
\textbf{Data availability} The data that support the findings of this study are available from the corresponding
authors upon reasonable request.

\bigskip
\textbf{Author contributions} 
B.K. designed the project. B.K. performed the LQSGW+DMFT calculations and conducted the data analysis. 
All authors wrote the manuscript, discussed the results, and commented on the paper.
\bibliography{ref}

\end{document}